\documentclass[12pt]{article}
\usepackage{amsmath,amsfonts,amssymb}  
\usepackage{graphicx}

 \ifx\pdfoutput\undefined
      \relax
 \else
      \usepackage{epstopdf}
 \fi

\makeatletter
\def\numberbysection{\@addtoreset{equation}{section}
    \def\theequation{\thesection.\arabic{equation}}}
\makeatother

\numberbysection

\newcommand{\be}{\begin{eqnarray}}
\newcommand{\ee}{\end{eqnarray}}
\newcommand{\non}{\nonumber}

\newcommand{\csch}{\mathop{\rm csch}\nolimits}
\newcommand{\sech}{\mathop{\rm sech}\nolimits}

\newcommand{\af}{\ensuremath{\mathsf{a}}}
\newcommand{\Af}{\ensuremath{\mathsf{A}}}

\newcommand{\Gf}{\ensuremath{\mathsf{G}}}
\newcommand{\Pf}{\ensuremath{\mathsf{P}}}

\def\ep{\epsilon}

\def\La{\Lambda}

\newcommand{\beq}{\begin{equation}}
\newcommand{\eeq}{\end{equation}}
\newcommand{\bea}{\begin{eqnarray*}}
\newcommand{\eea}{\end{eqnarray*}}
\newcommand{\beqa}{\begin{eqnarray}}
\newcommand{\eeqa}{\end{eqnarray}}

\begin{document}

\begin{titlepage}
\strut\hfill
\vspace{.5in}
\begin{center}

\LARGE NLIE and finite size effects of the spin-1/2 XXZ and \\ 
\LARGE sine-Gordon models with two boundaries revisited\\[1.0in]
\large Rajan Murgan\footnote{email: rmurgan@gustavus.edu}\\[0.8in]
\large Physics Department,\\ 
\large Gustavus Adolphus College,\\ 
\large 800 West College Avenue, St. Peter, MN 56082 USA\\
      
\end{center}

\vspace{.5in}

\begin{abstract}
Starting from the $T$-$Q$ equation of an open integrable spin-$\frac{1}{2}$ XXZ quantum spin chain with nondiagonal boundary terms,
we derive a nonlinear integral equation (NLIE) of the sine-Gordon model on a finite interval. We compute the boundary energy 
and the Casimir energy for the sine-Gordon model with both left and right boundaries. A relation between the boundary parameters
of the continuum model and the lattice model is given. We also present numerical results for the effective central charge of an open 
spin-$\frac{1}{2}$ XXZ quantum spin chain which find agreement with our analytical result for the central charge of the sine-Gordon model 
in the ultraviolet (UV) limit. 
  
\end{abstract}

\end{titlepage}

\setcounter{footnote}{0}

\section{Introduction}\label{sec:intro}

Due to applications in statistical mechanics and condensed matter physics, spin-$\frac{1}{2}$ XXZ quantum spin chain and sine-Gordon models with two boundaries have been 
subjected to intensive studies over the years \cite{Ga}-\cite{ABNPT}. Works on these topics have covered mainly diagonal and up to certain extent, nondiagonal boundary interactions. 
Proposal of Bethe ansatz solutions for the XXZ quantum spin chain with nondiagonal boundary terms \cite{nondiagonal, Nep} have made it possible to study these models further \cite{AN, ABNPT}.  
One such study is the investigation of the finite-size effects in sine-Gordon models using the nonlinear integral equation (NLIE) approach \cite{KBP, DdV}. Also refer to \cite{LMSS, Ra}
for further work on the subject.    

Motivated by previous works on the subject, we use a recently proposed solution of an open spin-$\frac{1}{2}$ XXZ quantum spin chain \cite{MN} to compute finite-size effects to the lowest 
energy state of the sine-Gordon model with two boundaries. In contrast to previous works, our work involves a boundary parameter free of any constraint (describing
nondiagonal boundary terms). However, one limitation of this solution 
is that the bulk anisotropy parameter for the XXZ spin chain assumes special values, namely $i\mu$, where $\mu = {\pi\over \nu}$, $\nu = 3, 5, 7,\ldots$. In particular, we derive a NLIE for the lowest energy
state of the sine-Gordon model on a finite interval and compute the boundary and Casimir energies. Our motivation is two fold: While 
the lowest energy state for other studied (critical) spin-$\frac{1}{2}$ XXZ quantum spin chain models is described by a sea of real Bethe roots, the lowest energy state 
for the spin-$\frac{1}{2}$ XXZ chain model considered here is described by a sea of ``two-strings'', i.e., complex conjugate pairs of Bethe roots, which is rather a characteristic of spin-$1$ 
XXZ chain. We thus feel it is worthwhile and interesting to compute finite size corrections for such a spin-$\frac{1}{2}$ model. Moreover, such a computation will serve as a 
useful guide when one considers the corresponding open spin-$1$ XXZ chain which has been associated with supersymmetric sine-Gordon models \cite{ssg1}-\cite{ANS}. As pointed out in \cite{ANS}, due to
the sea of ``two-strings'', familiar method of deriving the NLIE based on Bethe ansatz equations and the counting function \cite{DdV} does not seem to work. Fortunately, an NLIE 
can still be derived from the model's $T-Q$ equation. Refer to \cite{Su1} for more details.

The outline of this article is as follows.  In section 2, we briefly review the sine-Gordon model and the Hamiltonian of the open spin-$\frac{1}{2}$ XXZ quantum spin chain with two boundaries \cite{AN, ABNPT}. 
In section 3, as a warm up exercise, we rederive the NLIE given in \cite{AN} for the open spin-$\frac{1}{2}$ XXZ/sine-Gordon model where the boundary parameters obey a certain constraint. However, in contrast
to the approach \cite{DdV} used in \cite{AN, ABNPT}, we employ a method that utilizes the $T-Q$ equation \cite{Su1} of the open spin-$\frac{1}{2}$ XXZ quantum spin chain model. In section 4, we give main 
results of the paper. We derive an NLIE for the sine-Gordon model again using the $T-Q$ equation for an open spin-$\frac{1}{2}$ XXZ quantum spin chain with nondiagonal boundary terms, but now one of the boundary 
parameters free of any constraint \cite{MN}. We also derive the boundary energy and Casimir energy for the lowest energy state for this case. Finally, we show that the analytical result derived for the 
central charge of the sine-Gordon model in the ultraviolet (UV) limit agrees with the numerical results for the central charge of an open spin-$\frac{1}{2}$ XXZ spin chain that we obtain by 
solving the XXZ spin chain model numerically for few finite number of sites $N$, and extrapolating the results to $N\rightarrow\infty$ limit using an algorithm due to 
Vanden Broeck and Schwartz \cite{VBS}-\cite{HS}. Also see \cite{ABBBQ}, where such an extrapolation technique was used to study certain properties of statistical mechanical systems. 
This is followed by a brief discussion of our results and some open problems in section 5.    
   
\section{The sine-Gordon model and open spin-$1/2$ XXZ spin chain with two boundaries}\label{sec:sG}

In this section, we briefly review the sine-Gordon model (reproduced from \cite{AN}) and the Hamiltonian of the open spin-$\frac{1}{2}$ XXZ quantum spin chain with two boundaries. 
The sine-Gordon model on the finite ``spatial'' interval $x \in \left[ x_{-} \,, x_{+} \right]$ is described by the (Euclidean) action
\be
{\cal S} = \int_{-\infty}^{\infty}dy 
\int_{x_{-}}^{x_{+}}dx\  {\cal A}(\varphi \,, \partial_{\mu} \varphi) 
+ \int_{-\infty}^{\infty}dy \left[ 
{\cal B}_{-}(\varphi \,, {d\varphi\over dy} )\Big\vert_{x=x_{-}} +
{\cal B}_{+}(\varphi \,, {d\varphi\over dy} )\Big\vert_{x=x_{+}} \right] \,,
\label{SGaction}
\ee 
where the bulk action is given by 
\be 
{\cal A}(\varphi \,, \partial_{\mu} \varphi) = 
{1\over 2}(\partial_{\mu} \varphi)^{2}
+ \mu_{bulk} \cos (\beta \varphi) \,,
\label{SGbulkaction}
\ee 
and the boundary action is given by 
\be
{\cal B}_{\pm}(\varphi \,, {d\varphi\over dy} ) = 
\mu_{\pm} \cos( {\beta\over 2} (\varphi - \varphi_{0}^{\pm}))
\pm {\pi\gamma_{\pm}\over \beta} {d\varphi\over dy}  \,.
\label{SGboundaction}
\ee
As noted in \cite{AN, ABNPT}, the action is similar to the one considered by Ghoshal and Zamolodchikov \cite{GZ}, except for two boundaries
instead of one. Moreover, the presence of an additional term depending on the ``time'' derivative of the field in the boundary action (\ref{SGboundaction}) is also well noted.
As mentioned in \cite{AN}, while such a term can be eliminated in the one-boundary case by including a term proportional to $\partial_{x} \partial_{y} \varphi$ in (\ref{SGbulkaction}), 
such a step would simply result in the elimination of only one of the two $\gamma_{\pm}$ parameters (say, $\gamma_{+}$), and in a shift of the other ($\gamma_{-} \mapsto \gamma_{-} - \gamma_{+}$)
in the two-boundary case. The continuum bulk coupling constant $\beta$ is related to the lattice bulk coupling constant $\mu$ by $\beta^{2} = 8(\pi - \mu) = 8 \pi (\nu - 1)/\nu$, taking 
$\nu = \pi/\mu$. 

In subsequent sections that follow, we shall consider the energy of the ground state of this model as a function of the interval length $L \equiv x_{+} - x_{-}$, for large $L$.  
The leading contribution which is of order $L$ and which does not depend on the boundary interactions is well known \cite{AZ2}.  The boundary correction of order $1$ is also known 
\cite{AZ1, BPT}.  In this paper, we shall mainly concentrate on the compution of the Casimir correction of order $1/L$, and derive a nonlinear integral equation \cite{KBP}-\cite{Ra} 
for the lowest energy state.  The length $L$ and the soliton mass $m$ (whose relation to $\mu_{bulk}$ is known \cite{AZ2}) are given by
\be
L = N \Delta \,, \qquad m={2\over \Delta} e^{-{\pi \Lambda\over \mu}} \,,
\label{continuumlimit}
\ee
respectively. In (\ref{continuumlimit}), $\Delta$ is the lattice spacing, which in the continuum limit, taken to be $\Delta\rightarrow 0$ together with $\Lambda\rightarrow \infty$ and $N\rightarrow \infty$ for 
the inhomogeneity parameter and number of lattice sites respectively. We shall return to this in following sections where the NLIE are derived.  
Further, as given in \cite{AN}, the boundary parameters in the continuum action ($\mu_{\pm} \,, \varphi_{0}^{\pm}$) is related to the boundary parameters of the lattice model 
($\alpha_{\pm} \,, \beta_{\pm}$) (that appears in the Hamiltonian of the open spin-$\frac{1}{2}$ XXZ quantum spin chain (see (\ref{Hamiltonian}) below)) by
\be
\sinh (\alpha_{\pm} + \beta_{\pm})
&=& {\mu_{\pm}\over \mu_{c}}
i e^{- {i\over 2}\beta \varphi_{0}^{\pm}} \,, \non \\
\sinh (\alpha_{\pm} - \beta_{\pm})
&=& {\mu_{\pm}\over \mu_{c}} 
i e^{+ {i\over 2}\beta \varphi_{0}^{\pm}} \,.
\label{boundparamreltn2}
\ee
where 
\be
\mu_{c} = \sqrt{2 \mu_{bulk}\over
\sin \left({\beta^{2}\over 8\pi}\right)} \,.
\label{muc}
\ee 
The Hamiltonian of the open spin-$\frac{1}{2}$ XXZ quantum spin chain is given by \cite{dVGR, GZ}
\be
{\cal H }&=& {1\over 2}\Big\{ \sum_{n=1}^{N-1}\left( 
\sigma_{n}^{x}\sigma_{n+1}^{x}+\sigma_{n}^{y}\sigma_{n+1}^{y}
+\cosh \eta\ \sigma_{n}^{z}\sigma_{n+1}^{z}\right)\non \\
&+&\sinh \eta \Big[ 
\coth \alpha_{-} \tanh \beta_{-}\sigma_{1}^{z}
+ \csch \alpha_{-} \sech \beta_{-}\big( 
\cosh \theta_{-}\sigma_{1}^{x} 
+ i\sinh \theta_{-}\sigma_{1}^{y} \big) \non \\
&-& \coth \alpha_{+} \tanh \beta_{+} \sigma_{N}^{z}
+ \csch \alpha_{+} \sech \beta_{+}\big( 
\cosh \theta_{+}\sigma_{N}^{x}
+ i\sinh \theta_{+}\sigma_{N}^{y} \big)
\Big] \Big\} \,,
\label{Hamiltonian}
\ee
where $\sigma^{x} \,, \sigma^{y} \,, \sigma^{z}$ are the usual Pauli
matrices, $\eta$ is the bulk anisotropy parameter, $\alpha_{\pm} \,,
\beta_{\pm} \,, \theta_{\pm}$ are the boundary parameters, and $N$
is the number of spins. 

We remark that (\ref{boundparamreltn2}) was derived for the case where the lattice parameters ($\alpha_{\pm} \,, \beta_{\pm}, \theta_{\pm}$) obey the following constraint
\be
\alpha_{-} + \beta_{-} + \alpha_{+} + \beta_{+} = \pm (\theta_{-} - 
\theta_{+}) + \eta k \,,
\label{constraint}
\ee
where $k$ is an even integer if $N$ is odd, and is an odd integer if $N$ is even. A convenient redefinition of bulk and boundary parameters can also be adopted \cite{AN}: 
\be
\eta = i \mu \,, \qquad 
\alpha_{\pm} = i \mu a_{\pm} \,, \qquad \beta_{\pm} = \mu b_{\pm} \,,
\qquad \theta_{\pm} = i \mu c_{\pm}
\label{newparams} \,,
\ee 
where $\mu\,, a_{\pm}\,, b_{\pm}\,, c_{\pm}$ are all real, 
with $0 < \mu < \pi$.
With the above redefinitions, the constraint relation (\ref{constraint}) yields the following pair of real contraints: 
\be
a_{-} + a_{+} &=& \pm |c_{-} - c_{+}| + k \,, \non \\
b_{-} + b_{+} &=& 0 \,.
\label{realconstraints}
\ee
In this paper, we shall consider only even $N$ case. We also remark (see \cite{AN} for details) that the 
lattice parameters $\theta_{\pm}$ are related to the continuum parameters $\gamma_{\pm}$ which appear in (\ref{SGboundaction}). We shall see that for the case studied in section 4, 
(\ref{boundparamreltn2}) still holds true. 

\section{NLIE of the spin-${1\over 2}$ XXZ/sine-Gordon with constraint nondiagonal boundary terms }\label{sec:XXZcons}

In this section, we shall rederive the NLIE for the spin-$\frac{1}{2}$ XXZ/sine-Gordon model 
(along with the order 1 and order $1/L$ correction to the energy)
with constraint (\ref{constraint}) among the lattice boundary parameters. In contrast
to the familiar approach \cite{DdV} used in \cite{AN}, we utilize  instead the model's $T-Q$ equation (describing the transfer-matrix eigenvalues $T(u)$)  and its analyticity properties. 
We shall follow closely steps utilized in \cite{ANS} and employ similar notations.

\subsection{$T-Q$ equation and NLIE}

The $T-Q$ equation of the inhomogeneous open spin-$\frac{1}{2}$ XXZ chain with general boundary conditions (but with the boundary parameters 
($\alpha_{\pm} \,, \beta_{\pm}, \theta_{\pm}$) obeying constraint (\ref{constraint})) is given by \cite{Nep}
\be 
T(u)&=& \sinh(2u + i \mu)\, \tilde{B}^{(+)}(u)\, \phi(u+\frac{i\mu}{2})
\frac{Q(u- i\mu)}{Q(u)} \non \\
&+& \sinh(2u - i \mu)\, \tilde{B}^{(-)}(u)\, \phi(u-\frac{i\mu}{2})
\frac{Q(u+ i\mu)}{Q(u)} \,,
\label{TQspin12}
\ee
where 
\be
\phi(u)&=& \sinh^N(u-\La) \sinh^N(u+\La)  \,, \non \\
\tilde{B}^{(\pm)}(u)&=&\sinh (u\pm \frac{i\mu A_{-}}{2})
	     \sinh (u\pm \frac{i\mu A_{+}}{2}) \cosh (u\mp \frac{i\mu B_{-}}{2})\cosh (u\mp \frac{i\mu B_{+}}{2})  \,, \non \\
Q(u)&=&\prod_{k=1}^{M}\sinh(u-v_k) \sinh(u+v_k) \,.
\ee
where the bulk anisotropy parameter is $\eta = i\mu$, and we have redefined the boundary parameters as $A_{\pm} = 2a_{\pm} - 1\,, B_{\pm} = 2ib_{\pm} + 1$. 
$\La$ is the inhomogeneity parameter which provides a mass scale (see (\ref{continuumlimit})). $v_k$ represents the Bethe roots which are also zeros of $Q(u)$. We note here that the $Q(u)$ above 
differs from that given in \cite{Nep} by a mere shift of ${\eta\over 2}$ but otherwise equivalent.

Next, following \cite{ANS}, we define the auxiliary functions $a(u)$ and $\bar a(u)$ by
\beq
a(u)={\sinh(2u+i\mu)\, \tilde{B}^{(+)}(u)\, \phi(u+\frac{i\mu}{2})\, Q(u-i\mu)\over{
      \sinh(2u-i\mu)\, \tilde{B}^{(-)}(u)\, \phi(u-\frac{i\mu}{2})\, Q(u+i\mu)}} \,, \qquad
      {\bar a}(u)=a(-u)={1\over a(u)}
\,.
\label{defa}
\eeq
The transfer-matrix eigenvalues then simply become
\be
T(u) &=& \sinh(2u-i\mu)\, 
\tilde{B}^{(-)}(u)\, \phi(u-\frac{i\mu}{2}){Q(u+i\mu)\over{Q(u)}}(1+a(u)) \non \\
&=&\sinh(2u+i\mu)\, 
\tilde{B}^{(+)}(u)\, \phi(u+\frac{i\mu}{2}){Q(u-i\mu)\over{Q(u)}}(1+{\bar 
a}(u)) \,,
\label{TQhalf}
\ee
Note that $T(u)$ does not have zeros near the real axis except for one simple zero on real axis at $u = 0$.
The Bethe Ansatz equations can be written as \cite{Nep}
\be
a(v_{k}) = -1 \,, \qquad k = 1\,, \ldots \,, M \,.
\ee 
We consider the lowest energy state with ${N\over 2}$ real Bethe roots, namely $M = {N\over 2}$. We shall only consider ``massless regime'' (with purely imaginary bulk anisotropy parameter),
$\eta = i\mu$, with $0<\mu<\pi$.
The regions in parameter space $A_{\pm}$ which yield real Bethe roots for the lowest energy state can be divided in the following way. (See \cite{AN} for further details on some numerical 
results about these parameter regions.)
\be
\begin{array}{r@{\ : \quad}l}
    I   & 0 < A_{\pm} < {2\pi\over \mu} \\
    II  & 0 < A_{+} < {2\pi\over \mu} \quad \& \quad 
    -{2\pi\over \mu}< A_{-} < -1 \\
    III &  -{2\pi\over \mu}< A_{\pm} < -1 \\
    IV  & -{2\pi\over \mu} < A_{+} < -1 \quad \& \quad 
    0< A_{-} < {2\pi\over \mu}
  \end{array}
  \label{regions}
\ee
In addition, due to (\ref{realconstraints}), the parameters $B_{\pm}$ satisfy 
\beq
B_{+} + B_{-} = 2
\label{consforB}
\eeq
Also as performed in \cite{ANS}, one can remove the root of $T(u)$ at the origin (as pointed out earlier, $T(u)$ does not have zeros near the real axis except for a simple zero at the origin)
by defining
\beq
\check{T}(u)={T(u)\over \mu(u)} \,, 
\eeq
where $\mu(u)$ is any function whose only real root is a simple zero at the origin,
that is $\mu(0)=0\,, \ \mu'(0)\ne 0$, where the prime denotes differentiation with respect to $u$. Hence, the new $T-Q$ equation becomes
\be
\check{T}(u) = t_{-}(u)\, {Q(u+i\mu)\over{Q(u)}}(1+a(u)) = 
t_{+}(u)\, {Q(u-i\mu)\over{Q(u)}} (1+\bar a(u)) \,,
\label{modifiedTQ}
\ee
where
\be
t_{\pm}(u)={\sinh(2u \pm i\mu )\over \mu(u)}\tilde{B}^{(\pm)}(u)\, 
\phi(u \pm \frac{i\mu}{2}) \,.
\label{tpm}
\ee 

Utilizing the analyticity of $\ln \check{T}(u)$ near the real axis, we have the following from Cauchy's theorem,
\beq
0=\oint_C du\ [\ln\check{T}(u)]'' e^{iku} \,,
\label{Cauchy}
\eeq
where the contour $C$ is chosen as in Figure 1, $\epsilon$ being small and positive.

\setlength{\unitlength}{0.8cm}
\begin{picture}(6,4)(-8,-2)
\put(-2.5,0){\line(1,0){5}}
\put(0,-1.5){\line(0,1){3}}
\put(-1.0,0.3){\vector(-1,0){0.2}}
\put(-2.0,0.3){\line(1,0){4}}
\put(1.0,-0.3){\vector(1,0){0.2}}
\put(-2.0,-0.3){\line(1,0){4}}
\put(0.75,0.5){$C_{1}$}
\put(0.75,-0.8){$C_{2}$}
\put(2.0,-0.3){\line(0,1){0.6}}
\put(-2.0,-0.3){\line(0,1){0.6}}
\put(2.2,0.1){$i\epsilon$}
\put(2.2,-0.3){$-i\epsilon$}
\put(-2.0,-2.2){$\textrm{Figure\, 1: Integration\, contour}$}
\end{picture}

\noindent As a result, after using (\ref{modifiedTQ}), (\ref{Cauchy}) can thus be written as
\be
0 &=& \int_{C_1} du\ \left[ \ln t_{+}(u) \right]'' e^{iku}+
\int_{C_1} du\ \left\{ \ln \left[{Q(u-i\mu)\over Q(u)}\right] \right\}'' e^{iku}
+\int_{C_1} du\ \left[ \ln (1+{\bar a}(u)) \right]'' e^{iku} \non\\
&+& \int_{C_2} du\ \left[ \ln t_{-}(u) \right]'' e^{iku} 
+ \int_{C_2} du\ \left\{ \ln\left[{Q(u+i\mu)\over Q(u)}\right] 
\right\}'' e^{iku}
+\int_{C_2} du\ \left[ \ln (1+a(u))\right]'' e^{iku} \non\\ 
\label{Cauchy2}
\ee
Following \cite{ANS}, we define Fourier transforms along $C_{2}$ and $C_{1}$ as 
\be
\widehat{Lf''}(k)=\int_{C_2} du\ [\ln f(u)]'' e^{iku} \,, \qquad
\widehat{{\cal L}f''}(k)=\int_{C_1} du\ [\ln f(u)]'' e^{iku} \,,
\label{fouriertransfdef}
\ee
respectively, and after exploiting the periodicity \footnote{This is to make the imaginary part of the argument negative.}
\beq
Q(u)=Q(u-i\pi),\quad u\in C_1,\qquad{\rm and}\qquad
Q(u+i\mu)=Q(u+i\mu-i\pi),\quad u\in C_2 \,.
\label{Qperiodicity}
\eeq
we have the following
\be
\int_{C_1}du\ \left\{ \ln \left[{Q(u-i\mu)\over Q(u)}\right]
\right\}''e^{iku}&=&
-\widehat{LQ''}(k)\left(e^{-\mu k}-e^{-\pi k}\right) \,, \non \\
\int_{C_2}du\ \left\{ \ln \left[{Q(u+i\mu)\over Q(u)}\right] 
\right\}'' e^{iku}&=&
\widehat{LQ''}(k)\left[e^{(\mu-\pi)k}-1\right] \,,
\ee
Further, defining
\be
C(k) \equiv \int_{C_1}du \left[ \ln t_{+}(u) \right]'' e^{iku}
+\int_{C_2}du \left[ \ln t_{-}(u) \right]'' e^{iku} \,,
\label{Ck}
\ee
we obtain from (\ref{Cauchy2})
\be
C(k) + \widehat{LQ''}(k)\left[e^{(\mu-\pi)k}-1- e^{-\mu k}+e^{-\pi k}\right] + \widehat{LA''}(k)+\widehat{{\cal L}{\bar A}''}(k) = 0
\ee
which leads to the following 
\beq
\widehat{LQ''}(k)={e^{ \frac{\pi k}{2}}\over 4 \cosh( \frac{\mu k}{2})
\sinh\left((\pi-\mu)\frac{k}{2}\right)}
\left[
\widehat{LA''}(k)+\widehat{{\cal L}{\bar A}''}(k)+C(k)\right] \,.
\label{logQ}
\eeq
Note that the following definitions have been adopted,
\beq
A(u) = 1+a(u)\,, \quad \quad  \bar{A}(u) = 1+\bar{a}(u)
\eeq
Having found (\ref{logQ}), we proceed to derive the NLIE for the lattice sine-Gordon model. Taking the Fourier transform of $a(u)$ (see (\ref{defa})) along 
$C_{2}$, one arrives at
\be
\widehat{La''}(k)&=&\int_{C_2}du\ \left\{
\ln\left[{Q(u-i\mu)\over Q(u+i\mu)}\right]\right\}'' e^{iku} \non \\
&+& \int_{C_2}du\
\left\{ \ln\left[{\sinh(2u+i\mu)\, \tilde{B}^{(+)}(u)\, \phi(u+\frac{i\mu}{2})\over
\sinh(2u-i\mu)\, \tilde{B}^{(-)}(u)\, \phi(u-\frac{i\mu}{2})}\right] 
\right\}'' e^{iku}\,,
\ee
Using (\ref{Qperiodicity}), we obtain
\be
\widehat{La''}(k)&=&\widehat{LQ''}(k)\left[e^{-\mu 
k}-e^{(\mu-\pi)k}\right]\non \\
&+& \int_{C_2}du\
\left\{ \ln\left[{\sinh(2u+i\mu)\, \tilde{B}^{(+)}(u)\, \phi(u+\frac{i\mu}{2})\over
\sinh(2u-i\mu)\, \tilde{B}^{(-)}(u)\, \phi(u-\frac{i\mu}{2})}\right] 
\right\}'' e^{iku} \,.
\label{loga}
\ee
Inserting (\ref{logQ}) into (\ref{loga}) yields the required NLIE for the sine-Gordon model in Fourier space,
\beq
\widehat{La''}(k)= {\widehat G}(k)\left[\widehat{LA''}(k)+\widehat{{\cal L}{\bar 
A}''}(k)\right]+ C_{T}(k) \,, \label{nlie}
\eeq
where
\be
{\widehat G}(k) &=&{\sinh\left((\pi-2\mu)\frac{k}{2}\right)
\over 2\cosh (\frac{\mu k}{2})\sinh\left((\pi-\mu)\frac{k}{2}\right)} 
\,, \label{Gk} \\
C_{T}(k) &=&{\widehat G}(k)\, C(k)+\int_{C_2}du\
\left\{ \ln\left[{\sinh(2u+i\mu)\, \tilde{B}^{(+)}(u)\, \phi(u+\frac{i\mu}{2})\over
\sinh(2u-i\mu)\, \tilde{B}^{(-)}(u)\, \phi(u-\frac{i\mu}{2})}\right] 
\right\}'' e^{iku} \,. \label{CTdef}
\ee
The second term in (\ref{CTdef}) yields
\be
D(k)&\equiv&\int_{C_2}du\
\left\{ \ln\left[{\sinh(2u+i\mu)\, \tilde{B}^{(+)}(u)\, \phi(u+\frac{i\mu}{2})\over
\sinh(2u-i\mu)\, \tilde{B}^{(-)}(u)\, \phi(u-\frac{i\mu}{2})}\right] 
\right\}'' e^{iku} \non\\
&=&
 2\pi\psi(k)\Big\{
s_{-}e^{-\left({-\mu |A_{-}|\over{2}}+\pi\right)k}+s_{+}e^{-\left({-\mu |A_{+}|\over{2}}+\pi\right)k}
-s_{-}e^{-{\mu |A_{-}|\over{2}}k}-s_{+}e^{-{\mu |A_{+}|\over{2}}k} \non \\
&+&
e^{-\left(\mu B_{-}+\pi\right){k\over 2}}+e^{-\left(\mu B_{+}+\pi\right){k\over 2}}-e^{-\left(-\mu B_{-}+\pi\right){k\over 2}}-e^{-\left(-\mu B_{+}+\pi\right){k\over 2}}\non\\
&+&
N(e^{i\La k}+e^{-i\La k})\left[e^{\left({\mu\over{2}}-\pi\right)k}-e^{-{\mu\over{2}}k}
\right]
\Big\} 
+ 2\pi\psi_2(k)\left[e^{\left({\mu\over{2}}-{\pi\over{2}}\right)k}-e^{-{\mu\over{2}}k}
\right]
\label{D}
\ee
where $s_{\pm}\equiv$ sgn($A_{\pm}$),  $\psi(k) \equiv {k\over 1-e^{-\pi k}}$ and  $\psi_2(k) \equiv {k\over 1-e^{-{\pi k\over 2}}}$
and we have used the following identities (see also \cite{ANS}), 
\beq
\int_{C_2} {du\over 2\pi}  \left[ \ln \sinh(u-i \alpha)\right]'' e^{iku}= 
e^{-k(\alpha - n \pi)}  \psi(k) \,, 
\label{psi}
\eeq
where $n$ is an integer such that $0 < \Re e(\alpha - n \pi) < \pi$, 
and
\beq
\int_{C_2} {du\over 2\pi} \left[ \ln \sinh(2u)\right]'' e^{iku}
=\psi_2(k) \,.
\label{psi2}
\eeq
Further, from (\ref{tpm}), (\ref{Ck}) and using the fact that $\tilde{B}^{\pm}(u)$ and $\phi(u\pm{i\mu\over 2})$ are analytic and nonzero near the real axis (hence changing the contour integral 
on $C_{1}$ into $-C_{2}$), one arrives at the following result for $C(k)$,
\be
C(k) &=& -\int_{C_2}du\
\left\{ \ln\left[{\sinh(2u+i\mu)\, \tilde{B}^{(+)}(u)\, \phi(u+\frac{i\mu}{2})\over
\sinh(2u-i\mu)\, \tilde{B}^{(-)}(u)\, \phi(u-\frac{i\mu}{2})}\right] 
\right\}'' e^{iku}\non \\ 
&-& \oint_{C}du\
\left[\ln\mu(u)\right]''e^{iku}
\label{Ck2}
\ee
Note that the first term in (\ref{Ck2}) is simply the negative of (\ref{D}). The second term in (\ref{Ck2}) reduces to $-2\pi k$. (Refer to our earlier discussion on $\mu(u)$). 
Using (\ref{Gk})-(\ref{D}) and (\ref{Ck2}), we finally obtain the following for $C_{T}(k)$, 
\be
C_{T}(k) &=& -2\pi k \Bigg\{ {N\cos(\La k)\over \cosh({\mu k\over 2})} 
+{s_{+}\sinh\left((\mu |A_{+}|-\pi)\frac{k}{2}\right)
   +s_{-}\sinh\left((\mu |A_{-}|-\pi)\frac{k}{2}\right)
\over 2\cosh({\mu k\over 2}) \sinh\left((\mu-\pi)\frac{k}{2}\right)} 
\non \\
& & - {\left[\sinh\left(\frac{k}{2}\mu B_{-}\right)
   +\sinh\left(\frac{k}{2}\mu B_{+}\right)\right]
\over 2\cosh({\mu k\over 2}) \sinh\left((\mu-\pi)\frac{k}{2}\right)} + {\cosh({\mu k\over 4}) \sinh\left((2\mu-\pi)\frac{k}{4}\right)\over
  \cosh({\mu k\over 2}) \sinh\left((\mu-\pi)\frac{k}{4}\right)} 
\Bigg\} \,. 
\ee
Converting (\ref{nlie}) to coordinate space and integrating twice,
we obtain 
\be
\ln a(u) &=& 
\int_{-\infty}^{\infty}du'\ G(u-u'+i\epsilon) \ln (1+a(u'-i\epsilon)) - 
\int_{-\infty}^{\infty}du'\ G(u-u'-i\epsilon) \ln (1+\bar 
a(u'+i\epsilon))\non \\
&-& i 2N \tan^{-1}\left({\sinh \frac{\pi u}{\mu}\over 
\cosh \frac {\pi \La}{\mu} } \right) + i\, P_{bdry}(u) +i\pi \,,
\label{spinhalflatticeNLIE}
\ee
In (\ref{spinhalflatticeNLIE}), $G(u)$ is the Fourier transform of $\widehat G(k)$ (see (\ref{Gk}))
\be
G(u) = {1\over 2\pi} \int_{-\infty}^{\infty}dk\ e^{-i k u}\ \widehat 
G(k) \,,
\label{FTdefG}
\ee
and $P_{bdry}(u)$ is given by
\be
P_{bdry}(u) = \int_{0}^{u}du'\, R(u') = \frac{1}{2} 
\int_{-u}^{u}du'\, R(u')\,,
\label{spinhalfPbdry}
\ee
where $R(u)$ refers to the Fourier transform of $\hat R(k)$ which is given below,
\be
\hat R(k) &=& -2\pi \Bigg\{  
{s_{+}\sinh\left((\mu |A_{+}|-\pi)\frac{k}{2}\right)
   +s_{-}\sinh\left((\mu |A_{-}|-\pi)\frac{k}{2}\right)
\over 2\cosh({\mu k\over 2}) \sinh\left((\mu-\pi)\frac{k}{2}\right)} 
\non \\
& & - {\left[\sinh\left(\frac{k}{2}\mu B_{-}\right)
   +\sinh\left(\frac{k}{2}\mu B_{+}\right)\right]
\over 2\cosh({\mu k\over 2}) \sinh\left((\mu-\pi)\frac{k}{2}\right)} + {\cosh({\mu k\over 4}) \sinh\left((2\mu-\pi)\frac{k}{4}\right)\over
  \cosh({\mu k\over 2}) \sinh\left((\mu-\pi)\frac{k}{4}\right)} 
\Bigg\} \,.
\ee
The factor $i\pi$ (integration constant) in (\ref{spinhalflatticeNLIE}) is obtained by considering the $u\rightarrow\infty$ limit of (\ref{defa}) and (\ref{spinhalflatticeNLIE}). 
Further, proceeding as in \cite{ANS}, one obtains the correct factor. We have explicitly checked that this procedure yields the same integration constant for all possible combinations of  the 
boundary parameters, namely all four regions given in (\ref{regions}).

Next, taking the continuum limit ($\La \rightarrow \infty\,, N\rightarrow\infty\,, \Delta\rightarrow 0)$, the term 
$-i2N \tan^{-1}\left({\sinh \frac{\pi u}{\mu}\over 
\cosh \frac {\pi \La}{\mu}} \right)$ becomes $-i 2mL \sinh \theta$ after defining the renormalized rapidity $\theta$ as
\be 
\theta = \frac{\pi u}{\mu} \,.
\label{renormrapidity}
\ee 
Thus, (\ref{spinhalflatticeNLIE}) becomes
\be
\ln \af(\theta) &=& 
\int_{-\infty}^{\infty}d\theta'\ \Gf (\theta-\theta'+i\varepsilon) \ln 
(1+\af(\theta'-i\varepsilon)) - 
\int_{-\infty}^{\infty}d\theta'\ \Gf(\theta-\theta'-i\varepsilon) \ln (1+\bar 
\af(\theta'+i\varepsilon))\non \\
&-& i 2mL \sinh \theta + i\, \Pf_{bdry}(\theta) +i\pi\,,
\label{NLIE2}
\ee
where following definitions have been used,
\be
\varepsilon= \frac{\pi \epsilon}{\mu}\,, \quad \!
\af(\theta)=a(\frac{\mu \theta}{\pi})\,, \quad \!
\Pf_{bdry}(\theta)=P_{bdry}(\frac{\mu \theta}{\pi})\,, \quad \!
\Gf(\theta)=\frac{\mu}{\pi}G(\frac{\mu \theta}{\pi})\,, 
\label{mathfrakdefs}
\ee 
By making following identifications
\be
f(\theta) \equiv \ln (-\bar \af(\theta))\,, \qquad P_{bdry}^{AN}(\theta) \equiv -\Pf_{bdry}(\theta)
\ee
where $P_{bdry}^{AN}(\theta)$ is labelled as $P_{bdry}(\theta)$ in \cite{AN}, and defining instead,
\be 
\theta = \pi u \,, \qquad \Gf(\theta)=G(\frac{\theta}{\pi})\,, 
\label{renormrapidity2}
\ee 
one arrives at the result found in \cite{AN} (see equation (3.26) of the reference), hence confirming that the result (\ref{NLIE2}) matches that of 
Ahn and Nepomechie. 

\subsection{Boundary and Casimir energies}\label{subsec:spinhalfvacuumCasimir}

In this section, we compute the boundary correction (order 1) and the Casimir correction (order $1/L$). We begin by following the prescription of 
Reshetikhin and Saleur \cite{RS} (see also \cite{ANS}), according to which the energy for the inhomogeneous case 
($\Lambda \ne 0$) is given by 
\beq
E=-\frac{g}{\Delta}\left\{ \frac{d}{du}\ln T(u)\Bigg\vert_{u=\Lambda+\frac{i\mu}{2}}
-\frac{d}{du}\ln T(u)\Bigg\vert_{u=\Lambda-\frac{i\mu}{2}}\right\} 
\,, \label{energydef}
\eeq
where $g$ is given by 
\be
g=-\frac{i\mu}{4\pi} \,. \label{energynormalization}
\ee 
Using the fact that
\be 
\frac{d}{du}\ln T(u)\Bigg\vert_{u=\Lambda\pm\frac{i\mu}{2}}=
\frac{d}{du}\ln 
T^{(\pm)}(u)\Bigg\vert_{u=\Lambda\pm\frac{i\mu}{2}} \,
\label{deff2}
\ee
where
\be
T^{(\pm)}(u)&=& \sinh(2u \pm i \mu)\, \tilde{B}^{(\pm)}(u)\, \phi(u\pm\frac{i\mu}{2})
\frac{Q(u\mp i\mu)}{Q(u)}
\label{Tpm}
\ee
and 
\be
[\ln f(u)]' = \int \frac{dk}{2\pi}\ \widehat{Lf'}(k)\ e^{-ik u} \,, \qquad
u \in C_{2} \,,
\label{FTfact}
\ee 
(which in fact follows from (\ref{fouriertransfdef})) (\ref{energydef}) reduces to
\be
E&=&-\frac{g}{\Delta}\frac{d}{du}\left\{ \ln T^{(+)}(u+\frac{i\mu}{2})-
\ln T^{(-)}(u-\frac{i\mu}{2})\right\} \bigg\vert_{u=\Lambda} \non 
\\
&=&-\frac{g}{\Delta}\int \frac{dk}{2\pi} e^{-ik\Lambda}\left[
e^{\frac{\mu k}{2}}\widehat{LT^{(+)'}}(k)-e^{-\frac{\mu k}{2}}\widehat{LT^{(-)'}}(k)\right] \,,
\label{FTenergy}
\ee
Using (\ref{Tpm}), one finds the following
\be
e^{\frac{\mu k}{2}}\widehat{LT^{(+)'}}(k)&-&e^{-\frac{\mu k}{2}}\widehat{LT^{(-)'}}(k) \non \\
&=&
e^{{\mu k\over{2}}}
\widehat{L\tilde{B}^{(+)'}}(k)+e^{(\mu-\pi)k}\widehat{L\phi'}(k) 
+ e^{(\mu-\frac{\pi}{2})k} \frac{2\pi\psi_2(k)}{(-ik)}\non \\
&-& e^{-{\mu 
k\over{2}}}\widehat{L\tilde{B}^{(-)'}}(k)
-e^{-\mu k}\widehat{L\phi'}(k) 
- e^{-\mu k} \frac{2\pi\psi_2(k)}{(-ik)}\non \\
&+&4e^{-\frac{\pi k}{2}}\sinh((\pi-\mu)\frac{k}{2})
\widehat{LQ'}(k) \,,
\label{tpmdiff}
\ee
But first one need to determine the explicit form for $\widehat{LQ'}(k)$ which needed to be substituted in (\ref{tpmdiff}). After using (\ref{logQ}), (\ref{Ck2}) and 
$\widehat{LQ'}(k) = {1\over (-i k)}\widehat{LQ''}(k)$, we obtain
\be
\widehat{LQ'}(k) &=& {e^{ \frac{\pi k}{2}}\over 4 \cosh( \frac{\mu k}{2})
\sinh\left((\pi-\mu)\frac{k}{2}\right)}
\Big\{
\widehat{LA'}(k)+\widehat{{\cal L}{\bar 
A}'}(k)+\widehat{L\tilde{B}^{(-)'}}(k)-\widehat{L\tilde{B}^{(+)'}}(k) \non \\
&+&
\left[e^{-{\mu k\over{2}}}-e^{\left({\mu\over{2}}-\pi\right)k}\right]
\widehat{L\phi'}(k) 
- \left[-e^{-{\mu k\over{2}}}+e^{\left({\mu\over{2}}-\frac{\pi}{2}\right)k}\right] 
\frac{2\pi\psi_2(k)}{(-ik)}
-2\pi i \Big\}
\,,
\ee
This eventually leads to the following result for the energy,
\beq
E = E_{L} + E_{1} + E_{1/L}
\label{FTenergyint}
\eeq
where
\be
E_{L} = -\frac{g}{2\pi \Delta}\int_{-\infty}^{\infty} dk\ e^{-ik\Lambda}\tanh({\mu k\over{2}})\left[2e^{-{\pi k\over 2}}\cosh((\mu-{\pi\over{2}})k) \widehat{L\phi'}(k)\right]
\label{bulk} 
\ee
\be
E_{1} = -\frac{g}{2\pi \Delta}\int_{-\infty}^{\infty} dk\ e^{-ik\Lambda}\tanh({\mu k\over{2}})\bigg[&-&{2\pi i\over \sinh \frac{\mu k}{2}}
+ {2\pi i\cosh((\mu-\frac{\pi}{4})k)\over \sinh \frac{\pi k}{4}}\non \\
&+& e^{-{\mu k\over{2}}}\widehat{L\tilde{B}^{(-)'}}(k)
+e^{{\mu 
k\over{2}}}\widehat{L\tilde{B}^{(+)'}}(k)
\bigg] 
\label{bound}
\ee
\be
E_{1/L} = -\frac{g}{2\pi \Delta}\int_{-\infty}^{\infty} dk\ e^{-ik\Lambda}\frac{1}{\cosh{\mu k\over{2}}}\left[
\widehat{LA'}(k)+\widehat{{\cal L}{\bar A}'}(k)\right]
\label{Cas} 
\ee
As the subscripts suggest, equations (\ref{bulk}), (\ref{bound}) and (\ref{Cas}) refer to the bulk, boundary and Casimir energies respectively.
Now, we shall evaluate each of these terms explicitly. Using 
\be
\widehat{L\phi'}(k) = {\widehat{L\phi''}(k)\over -ik}
\label{FT}
\ee 
and the identity (\ref{psi}), we have the following,
\be
\widehat{L\phi'}(k)=2\pi N(e^{i\Lambda k}+e^{-i\Lambda 
k})\frac{\psi(k)}{(-ik)} \,.
\ee
which upon substitution in (\ref{bulk}), yields
\beq
E_{L}= \frac{Ng}{i \Delta}\int_{-\infty}^{\infty} dk\ (1+e^{-2i\Lambda k})
\frac{\sinh{\mu k\over{2}}\cosh\left((\mu-{\pi\over{2}})k\right)}
{\cosh{\mu k\over{2}}\sinh{\pi k\over{2}}} \,.
\label{Bulkenergy}
\eeq
Adopting the renormalization procedure \cite{DdV} of discarding divergent terms ($\Lambda$-independent term),
keeping only the finite terms, evaluating the remaining
integral by closing the contour in the lower half plane and selecting
only the contribution from the pole at $k=-\frac{i\pi}{\mu}$ and using (\ref{continuumlimit}) and (\ref{energynormalization}) in the process, 
we finally arrive at the result found in \cite{DdV, AZ2}, namely
\beq
E_{L}=\frac{1}{4}Lm^2\cot\frac{\pi^2}{2\mu} \,,
\eeq

We now consider the boundary energy (\ref{bound}). Using results (\ref{FT}) (with $\phi\rightarrow \tilde{B}^{(\pm)}$) and (\ref{psi}), 
we have
\be
e^{-{\mu k\over{2}}}\widehat{L\tilde{B}^{(-)'}}(k)+e^{{\mu 
k\over{2}}}\widehat{L\tilde{B}^{(+)'}}(k)
&=&\frac{2i\pi}{\sinh\frac{\pi k}{2}}\Big[
\cosh[\frac{\pi k}{2}-s_{-}(\frac{\mu k}{2}+\frac{\mu k}{2}A_{-})]\non \\
&+& \cosh[\frac{\pi k}{2}-s_{+}(\frac{\mu k}{2}+\frac{\mu k}{2}A_{+})]+\cosh(\frac{k\mu}{2}B_{-}-\frac{k\mu}{2})\non \\
&+& \cosh(\frac{k\mu}{2}B_{+}-\frac{k\mu}{2})\Big] 
\label{bpbmdiff}
\ee 
Substituting (\ref{bpbmdiff}) into (\ref{bound}), we obtain the boundary energy
\be
E_{1}={m\over 2}\left[1 + \cot \frac{\pi\nu}{4}-
 \frac{\cos\left(\frac{\nu}{2}(\pi-s_{-}\mu(1+A_{-}))\right)}{\sin\frac{\pi\nu}{2}}
 - \frac{\sin\left(\frac{B_{-}\pi}{2}\right)}{\sin\frac{\pi\nu}{2}}+(-\leftrightarrow +)\right] \,
\label{FTbenergy}
\ee
which is evaluated using the same contour as for the bulk energy. The symbol $(-\leftrightarrow +)$ represents the terms with 
$A_{-}\rightarrow A_{+}\,, B_{-}\rightarrow B_{+}\,, s_{-}\rightarrow s_{+}$.
In terms of $a_{\pm}$ and $b_{\pm}$, the above becomes
\be
E_{1}={m\over 2}\left[1 + \cot \frac{\pi\nu}{4}-
 \frac{\cos\left(\frac{\nu}{2}(\pi-2s_{-}\mu a_{-})\right)}{\sin\frac{\pi\nu}{2}}
 - \frac{\cosh\left(b_{-}\pi\right)}{\sin\frac{\pi\nu}{2}}+(-\leftrightarrow +)\right] \,
\label{FTbenergy2}
\ee
which is the expression found by Ahn and Nepomechie in \cite{AN}. 

Finally, we consider the Casimir energy given by (\ref{Cas}).
As with the NLIE, passing to coordinate space and taking the continuum limit, we obtain
\be
E_C&=&\frac{2g}{i\Delta\mu}
\int_{-\infty}^{\infty}du\,\Im m \left(\frac{1}{\cosh\frac{\pi}{\mu}
(\Lambda-u-i\ep)}\right)'\ln (1+{\bar a}(u+i\ep))
\ee
Further, $\left(\frac{1}{\cosh\frac{\pi}{\mu}(\Lambda-u-i\ep)}\right)'\rightarrow{2\pi\over \mu}e^{-{\pi\over \mu}(\Lambda-u-i\epsilon)}$ at 
$\Lambda\rightarrow\infty$ limit. Using (\ref{continuumlimit}), we have
\be
E_{C} = -\frac{m}{2\mu}
\int_{-\infty}^{\infty}du\, \Im m \, e^{\frac{\pi}{\mu}(u+i\ep)}
\ln (1+{\bar a}(u+i\ep))\,
\ee
From (\ref{renormrapidity}) and the first two definitions in (\ref{mathfrakdefs}) and after some manipulation, $E_{C}$ reduces to (Note also that we have used ${\bar \Af}(u)=\Af(-u)$ 
and $\Im m z=-\Im m{\bar z}$ in the process.), 
\be
E_{C} = -\frac{m}{2\pi}
\int_{-\infty}^{\infty}d\theta\, \Im m \, \sinh(\theta+i\varepsilon)\ln (1+{\bar \af}(\theta+i\varepsilon))\,
\ee
Invoking the identification $f(\theta) \equiv \ln(-{\bar \af}(\theta))$, the above becomes
\be
E_{C} = -{m\over 2\pi} \int_{-\infty}^{\infty} d\theta\ 
\Im m\ 
\sinh (\theta+ i \varepsilon)  \ln (1 - e^{f(\theta + i \varepsilon)})
\,.
\label{SGCasimir}
\ee
which is indeed the result derived in \cite{AN}.

\section{NLIE of the spin-${1\over 2}$ XXZ/sine-Gordon with nondiagonal boundary terms }\label{sec:XXZnocons}
In this section, we give the main results of the paper. We derive a NLIE for an open spin-$\frac{1}{2}$ XXZ/sine-Gordon model with the lattice boundary parameters 
satisfying the following,
\be
\alpha_{\pm} = \eta\, \qquad \beta_{+} = \beta_{-} = \beta\, \qquad \theta_{-} = \theta_{+} = 0 \,.
\label{bc}
\ee 
Note that the above parameters do not obey the constraint (\ref{constraint}). Moreover, we take $\beta$ to be arbitrary and real. The bulk anisotropy parameter $\eta$, equals $i\mu$, where $\mu={\pi\over \nu}\,, \nu=3,5,7,\ldots$. 
As in section 3, we will also derive the order 1 and $1/L$ corrections to the energy. We shall again use the spin-$\frac{1}{2}$ XXZ quantum spin chain model's $T-Q$ equation \cite{MN}.  
We then consider the UV limit of the central charge for the sine-Gordon model. We also present some numerical results.

\subsection{$T-Q$ equation and NLIE}

The $T-Q$ equation of the inhomogeneous open spin-$\frac{1}{2}$ XXZ chain with nondiagonal boundary conditions, with the boundary parameters 
($\alpha_{\pm} \,, \beta_{\pm}\,, \theta_{\pm}$) specified by (\ref{bc}) is given by \cite{MN},
\be 
T(u)&=& \sinh(2u + i \mu)\, \sinh(u + \frac{i\mu}{2})\, \sinh(u - \frac{3i\mu}{2})\, \mathcal{\tilde{B}}^{(+)}(u)\, \phi(u+\frac{i\mu}{2})
\frac{Q(u+ i\mu\,(\nu-1))}{Q(u)} \non \\
&+& \sinh(2u - i \mu)\, \sinh(u - \frac{i\mu}{2})\, \sinh(u + \frac{3i\mu}{2})\, \mathcal{\tilde{B}}^{(-)}(u)\, \phi(u-\frac{i\mu}{2})
\frac{Q(u- i\mu\,(\nu-1))}{Q(u)} \non\\
\label{TQspin12b}
\ee
where 
\be
\phi(u)&=& \sinh^N(u-\La) \sinh^N(u+\La)  \,, \non \\
\mathcal{\tilde{B}}^{(\pm)}(u)&=&\sinh\big(\frac{1}{2}(u\mp \frac{i\mu A_{+}}{2})\big)
	    \sinh\big(\frac{1}{2}(u\mp \frac{i\mu B_{+}}{2})\big) \cosh\big(\frac{1}{2}(u\mp \frac{i\mu A_{-}}{2})\big)\non \\
&\times& \cosh\big(\frac{1}{2}(u\mp \frac{i\mu B_{-}}{2})\big)  \,, \non \\
Q(u)&=&\prod_{k=1}^{M}\sinh\big(\frac{1}{2}(u-v_k-\frac{i\pi}{2})\big) \sinh\big(\frac{1}{2}(u+v_k+\frac{i\pi}{2})\big)  \,, \non \\
\nu &=& 3,5,7,\ldots\,, \qquad M = N+\nu-1\,.
\label{defs}
\ee
In (\ref{defs}), 
\be
A_{\pm} = 1+2ib_{\pm}-\nu \,, \qquad B_{\pm} = 1-2ib_{\pm}+\nu \,.
\ee
where $b_{\pm}$ is related to $\beta_{\pm}$ (that appear in spin chain Hamiltonian (\ref{Hamiltonian})) as in (\ref{newparams}). Note that $\beta_{-}=\beta_{+}=\beta$
implies $b_{+}=b_{-}=b$. This in turn yields
\be
A_{+} = A_{-} = 1 + 2ib - \nu \,, \qquad B_{+} = B_{-} = 1 - 2ib + \nu 
\ee
$\{v_{k}+\frac{i\pi}{2}\}$ represent the Bethe roots (and the zeros of $Q(u)$). The $N + \nu - 1$ ``shifted roots'' $\{v_{1}\,,\ldots\,,v_{N+\nu-1}\}$ for the lowest energy state have the following structure,
\be
 \left\{ \begin{array}{c@{\quad : \quad} l}
w_{k} \pm {i \pi\over 2} & k = 1\,, 2\,, \ldots \,, {N\over 2} \\
w_{k}^{(1)} + i \pi\,, \quad w_{k}^{(2)} & k = 1\,, 2\,, \ldots \,, {\nu-1\over 2}
\end{array} \right. \,,
\label{stringhypothesisI}
\ee 
where $\{ w_{k}\,, w_{k}^{(1)}\,, w_{k}^{(2)}\}$ are all {\em real} and
positive. $\big\{w_{k}\big\}$ are the sea-roots while $\big\{w_{k}^{(1)}\,, w_{k}^{(2)}\big\}$ represent the extra-roots, which are not part of the sea-roots. As pointed out earlier, an interesting feature of this model is 
that the lowest energy state is described by ${N\over 2}$ ``strings'' of length 2, in addition to ${\nu-1\over 2}$ pairs 
of ``strings'' of length 1. By choosing $\beta_{+} = \beta_{-}$,  we find that $w_{k}^{(1)} = w_{k}^{(2)}$. Numerical studies indicate that for $N\rightarrow\infty$, $w_{k}^{(1)}$ becomes large as well. It is also worth mentioning that from numerical results for the 
Bethe roots, $w_{j}^{(1)} > w_{k}$ for all possible values of $j$ and $k$, recalling the fact that $j=1\,, 2\,, \ldots \,, {\nu-1\over 2}$  and $k=1\,, 2\,, \ldots \,, {N\over 2}$.

Next, as in section 3, we define the corresponding auxiliary functions $a(u)$ and $\bar{a}(u)$ by
\be
a(u)&=&{\sinh(2u+i\mu)\, \sinh(u+{i\mu\over 2})\, \sinh(u-{3i\mu\over 2})\, \mathcal{\tilde{B}}^{(+)}(u)\, \phi(u+\frac{i\mu}{2})\, Q(u+i\mu(\nu-1))\over{
      \sinh(2u-i\mu)\, \sinh(u-{i\mu\over 2})\, \sinh(u+{3i\mu\over 2})\, \mathcal{\tilde{B}}^{(-)}(u)\, \phi(u-\frac{i\mu}{2})\, Q(u-i\mu(\nu-1))}} \non \\
      {\bar a}(u)&=&a(-u)={1\over a(u)}\,.
\label{defa2}
\ee
Using (\ref{defa2}), the $T-Q$ equation (\ref{TQspin12b}) becomes
\be
T(u) &=& \sinh(2u-i\mu)\, \sinh(u-{i\mu\over 2})\, \sinh(u+{3i\mu\over 2})\,
\mathcal{\tilde{B}}^{(-)}(u)\, \phi(u-\frac{i\mu}{2})\non \\
&\times& {Q(u-i\mu(\nu-1))\over{Q(u)}}(1+a(u)) \,, \non \\
&=&\sinh(2u+i\mu)\, \sinh(u+{i\mu\over 2})\, \sinh(u-{3i\mu\over 2})\,
\mathcal{\tilde{B}}^{(+)}(u)\, \phi(u+\frac{i\mu}{2})\non \\
&\times& {Q(u+i\mu(\nu-1))\over{Q(u)}}(1+{\bar a}(u)) \,,
\label{TQhalf2}
\ee
As before, $T(u)$ does not have zeros near the real axis except for one simple zero on real axis at $u = 0$.
The Bethe Ansatz equations follow from (\ref{TQhalf2}), 
\be
a(v_{k}+\frac{i\pi}{2}) = -1 \,, \qquad k = 1\,, \ldots \,, M=N+\nu-1 \,.
\label{BAE2}
\ee 
Following the steps of section 3, we remove the root of $T(u)$ at the origin by defining
\beq
\check{T}(u)={T(u)\over \mu(u)} \,, 
\eeq
where $\mu(u)$ is a function whose only real root is a simple zero at the origin,
with $\mu(0)=0\,, \ \mu'(0)\ne 0$. In terms of $\check{T}(u)$, the $T-Q$ equation becomes
\be
\check{T}(u) = t_{-}(u)\, {Q(u-i\mu(\nu-1))\over{Q(u)}}(1+a(u)) = 
t_{+}(u)\, {Q(u+i\mu(\nu-1))\over{Q(u)}} (1+\bar a(u)) \,,
\label{modifiedTQ2}
\ee
where
\be
t_{\pm}(u)={\sinh(2u \pm i\mu )\sinh(u\pm{i\mu\over 2})\sinh(u\mp{3i\mu\over 2})\over \mu(u)}\mathcal{\tilde{B}}^{(\pm)}(u)\, 
\phi(u \pm \frac{i\mu}{2}) \,.
\label{tpm2}
\ee 

Exploiting the analyticity of $\ln \check{T}(u)$ near the real axis, we have 
\be
0&=&\oint_C du\ [\ln\check{T}(u)]'' e^{iku} \non \\
&=& \int_{C_1} du\ \left[ \ln t_{+}(u) \right]'' e^{iku}+
\int_{C_1} du\ \left\{ \ln \left[{Q(u+i\mu(\nu-1))\over Q(u)}\right] \right\}'' e^{iku}
+\int_{C_1} du\ \left[ \ln (1+{\bar a}(u)) \right]'' e^{iku} \non\\
&+& \int_{C_2} du\ \left[ \ln t_{-}(u) \right]'' e^{iku} 
+ \int_{C_2} du\ \left\{ \ln\left[{Q(u-i\mu(\nu-1))\over Q(u)}\right] 
\right\}'' e^{iku}
+\int_{C_2} du\ \left[ \ln (1+a(u))\right]'' e^{iku} \non\\ 
\label{Cauchy3}
\ee
where the contour $C$ is again chosen as in Figure 1.

We first carefully separate the extra-roots terms $\big\{w_{k}^{(1)}\big\}$ in $Q(u)$ before proceeding with further computation involving Fourier transforms 
defined in (\ref{fouriertransfdef}). Introducing (\ref{stringhypothesisI}) in (\ref{defs}), we get
\be
Q(u) &=& \prod_{k=1}^{{N\over 2}}{1\over 4}\sinh(u-w_{k})\sinh(u+w_{k})\non\\
&\times& \prod_{k=1}^{{\nu-1\over 2}}\sinh\big({1\over 2}(u-w_{k}^{(1)}-{i\pi\over 2})\big)\sinh\big({1\over 2}(u+w_{k}^{(1)}+{i\pi\over 2})\big)\non \\
&\times& \sinh\big({1\over 2}(u-w_{k}^{(1)}+{i\pi\over 2})\big)\sinh\big({1\over 2}(u+w_{k}^{(1)}-{i\pi\over 2})\big)\,.
\label{Qex}
\ee
Defining
\be
\tilde{Q}(u) &=& \prod_{k=1}^{{N\over 2}}{1\over 4}\sinh(u-w_{k})\sinh(u+w_{k})
\ee
and using (\ref{fouriertransfdef}) and (\ref{psi}), we arrive at the following result for the Fourier transform of $[\ln Q(u)]''$,
\be
\widehat{LQ''}(k) = \widehat{L\tilde{Q}''}(k) + {4\pi k\over \sinh(\pi k)}\sum_{j=1}^{{\nu-1\over 2}}\cosh({k\pi\over 2})\cos(k\,w_{j}^{(1)})\,. 
\ee
Using the periodicity for $\tilde{Q}(u)$, namely
\beq
\tilde{Q}(u)=\tilde{Q}(u-i\pi),\quad u\in C_1,\qquad{\rm and}\qquad
\tilde{Q}(u+i\mu)=\tilde{Q}(u+i\mu-i\pi),\quad u\in C_2 \,
\label{Qperiodicity2}
\eeq
and the identity (\ref{psi}), the second and the fifth terms in (\ref{Cauchy3}) reduce to
\be
\int_{C_1}du\ \left\{ \ln \left[{Q(u+i\mu(\nu-1))\over Q(u)}\right]
\right\}''e^{iku} &+& \int_{C_2}du\ \left\{ \ln \left[{Q(u-i\mu(\nu-1))\over Q(u)}\right] 
\right\}'' e^{iku} \non \\
&=& -4e^{-{\pi k\over 2}}\cosh({\mu k\over 2})\sinh\big({k\over 2}(\pi-\mu)\big)\widehat{L\tilde{Q}''}(k)\non \\
&+& {8\pi k\over \sinh(\pi k)}\sinh(k\mu)\sum_{j=1}^{{\nu-1\over 2}}\cosh({k\pi\over 2})\cos(k\,w_{j}^{(1)})\non \\ 
\ee
As in section 3, we define
\be
C(k) \equiv \int_{C_1}du \left[ \ln t_{+}(u) \right]'' e^{iku}
+\int_{C_2}du \left[ \ln t_{-}(u) \right]'' e^{iku} \,.
\label{Ck3}
\ee
As a result, we obtain the following from (\ref{Cauchy3}),
\be
\widehat{L\tilde{Q}''}(k)&=&{e^{ \frac{\pi k}{2}}\over 4 \cosh( \frac{\mu k}{2})
\sinh\left((\pi-\mu)\frac{k}{2}\right)}
\left[\widehat{LA''}(k)+\widehat{{\cal L}{\bar A}''}(k)+C(k)\right] \non \\
&+& {4\pi k \sinh({k\mu\over 2})e^{ \frac{\pi k}{2}}\over \sinh(\pi k)\sinh\left((\pi-\mu)\frac{k}{2}\right)}\sum_{j=1}^{{\nu-1\over 2}}\cosh({k\pi\over 2})\cos(k\,w_{j}^{(1)})\,.
\label{logQ2}
\ee
Following definitions have again been adopted,
\beq
A(u) = 1+a(u)\,, \quad \quad  \bar{A}(u) = 1+\bar{a}(u)
\eeq
Next, we proceed to derive the NLIE for the lattice sine-Gordon model. Fourier transform of $a(u)$ (see (\ref{defa2})) along 
$C_{2}$ yields,
\be
\widehat{La''}(k)&=&\int_{C_2}du\ \left\{
\ln\left[{Q(u+i\mu(\nu-1))\over Q(u-i\mu(\nu-1))}\right]\right\}'' e^{iku} \non \\
&+& \int_{C_2}du\
\left\{ \ln\left[{\sinh(2u+i\mu)\, \sinh(u+{i\mu\over 2})\, \sinh(u-{3i\mu\over 2})\, \mathcal{\tilde{B}}^{(+)}(u)\, \phi(u+\frac{i\mu}{2})\over
\sinh(2u-i\mu)\, \sinh(u-{i\mu\over 2})\, \sinh(u+{3i\mu\over 2})\, \mathcal{\tilde{B}}^{(-)}(u)\, \phi(u-\frac{i\mu}{2})}\right] 
\right\}'' e^{iku}\,. \non \\
\ee
Using (\ref{Qperiodicity2}) and (\ref{psi}), we obtain
\be
\widehat{La''}(k)&=&\widehat{L\tilde{Q}''}(k)\left[e^{-\mu 
k}-e^{(\mu-\pi)k}\right]\non \\
&-& {8\pi k\over \sinh(\pi k)}\sinh(k\mu)\sum_{j=1}^{{\nu-1\over 2}}\cosh({k\pi\over 2})\cos(k\,w_{j}^{(1)})\non \\  
&+& \int_{C_2}du\
\left\{ \ln\left[{\sinh(2u+i\mu)\, \sinh(u+{i\mu\over 2})\, \sinh(u-{3i\mu\over 2})\, \mathcal{\tilde{B}}^{(+)}(u)\, \phi(u+\frac{i\mu}{2})\over
\sinh(2u-i\mu)\, \sinh(u-{i\mu\over 2})\, \sinh(u+{3i\mu\over 2})\, \mathcal{\tilde{B}}^{(-)}(u)\, \phi(u-\frac{i\mu}{2})}\right] 
\right\}'' e^{iku} \,.\non \\
\label{loga2}
\ee
Inserting (\ref{logQ2}) into (\ref{loga2}) yields the following NLIE for the sine-Gordon model in Fourier space,
\be
\widehat{La''}(k) &=& {\widehat G}(k)\left[\widehat{LA''}(k)+\widehat{{\cal L}{\bar A}''}(k)\right]+ C_{T}(k)\non \\ 
&-& {4\pi k\over \cosh({\pi k\over 2})\sinh\big({k\over 2}(\pi-\mu)\big)}\sinh({k\mu\over 2})\sum_{j=1}^{{\nu-1\over 2}}\cosh({k\pi\over 2})\cos(k\,w_{j}^{(1)})\,,\non \\ 
\label{nlie2b}
\ee
where
\be
{\widehat G}(k) &=&{\sinh\left((\pi-2\mu)\frac{k}{2}\right)
\over 2\cosh (\frac{\mu k}{2})\sinh\left((\pi-\mu)\frac{k}{2}\right)} 
\,, \label{Gk2b} \\
C_{T}(k) &=&{\widehat G}(k)\, C(k)\non \\
&+& \int_{C_2}du\
\left\{ \ln\left[{\sinh(2u+i\mu)\, \sinh(u+{i\mu\over 2})\, \sinh(u-{3i\mu\over 2})\, \mathcal{\tilde{B}}^{(+)}(u)\, \phi(u+\frac{i\mu}{2})\over
\sinh(2u-i\mu)\, \sinh(u-{i\mu\over 2})\, \sinh(u+{3i\mu\over 2})\, \mathcal{\tilde{B}}^{(-)}(u)\, \phi(u-\frac{i\mu}{2})}\right] 
\right\}'' e^{iku} \,. \non \\
\label{CTdef2}
\ee

Adopting similar steps of the previous section, and using the following for $C(k)$ (which is obtained from (\ref{Ck3})), 
\be
C(k) &=& -\int_{C_2}du\
\left\{ \ln\left[{\sinh(2u+i\mu)\, \sinh(u+{i\mu\over 2})\, \sinh(u-{3i\mu\over 2})\, \mathcal{\tilde{B}}^{(+)}(u)\, \phi(u+\frac{i\mu}{2})\over
\sinh(2u-i\mu)\, \sinh(u-{i\mu\over 2})\, \sinh(u+{3i\mu\over 2})\, \mathcal{\tilde{B}}^{(-)}(u)\, \phi(u-\frac{i\mu}{2})}\right] 
\right\}'' e^{iku}\non \\ 
&-& \oint_{C}du\
\left[\ln\mu(u)\right]''e^{iku}
\label{Ck2bb}
\ee
we find, 
\be
C_{T}(k) &=& -2\pi k \Bigg\{ {N\cos(\La k)\over \cosh({\mu k\over 2})} 
- {\sinh({\mu k\over 2}) \cosh\left((2\mu-\pi)\frac{k}{2}\right)\over
  \cosh({\mu k\over 2}) \sinh\left((\mu-\pi)\frac{k}{2}\right)}\non \\
& & + {\cosh({\mu k\over 4}) \sinh\left((2\mu-\pi)\frac{k}{4}\right)\over
  \cosh({\mu k\over 2}) \sinh\left((\mu-\pi)\frac{k}{4}\right)}
- {\sinh\left(\frac{k}{2}\mu\right)\cos(k\mu b)
\over \cosh({\mu k\over 2}) \sinh\left((\mu-\pi)\frac{k}{2}\right)}\non \\
& & +{2\, \sinh({k\mu\over 2})\over \cosh({\pi k\over 2})\sinh\big((\pi-\mu){k\over 2}\big)}\sum_{j=1}^{{\nu-1\over 2}}\cosh({k\pi\over 2})\cos(k\,w_{j}^{(1)})
\Bigg\} \,. 
\ee
The second term in (\ref{Ck2bb}) reduces to $-2\pi k$.
Converting (\ref{nlie2b}) to coordinate space and integrating twice,
we obtain 
\be
\ln a(u) &=& 
\int_{-\infty}^{\infty}du'\ G(u-u'+i\epsilon) \ln (1+a(u'-i\epsilon)) - 
\int_{-\infty}^{\infty}du'\ G(u-u'-i\epsilon) \ln (1+\bar 
a(u'+i\epsilon))\non \\
&-& i 2N \tan^{-1}\left({\sinh \frac{\pi u}{\mu}\over 
\cosh \frac {\pi \La}{\mu} } \right) + i\, P_{bdry}(u) +{i\pi\over \nu-1} \,,
\label{spinhalflatticeNLIE2}
\ee
where we recall that $G(u)$ is the Fourier transform of $\widehat G(k)$ (see (\ref{Gk2b}))
\be
G(u) = {1\over 2\pi} \int_{-\infty}^{\infty}dk\ e^{-i k u}\ \widehat 
G(k) \,,
\label{FTdefG2}
\ee
and $P_{bdry}(u)$ is given by
\be
P_{bdry}(u) = \int_{0}^{u}du'\, R(u') = \frac{1}{2} 
\int_{-u}^{u}du'\, R(u')\,,
\label{spinhalfPbdry2}
\ee
where $R(u)$ refers to the Fourier transform of $\hat R(k)$ given below,
\be
\hat R(k) &=& -2\pi \Bigg\{ 
- {\sinh({\mu k\over 2}) \cosh\left((2\mu-\pi)\frac{k}{2}\right)\over
  \cosh({\mu k\over 2}) \sinh\left((\mu-\pi)\frac{k}{2}\right)}\non \\
& & + {\cosh({\mu k\over 4}) \sinh\left((2\mu-\pi)\frac{k}{4}\right)\over
  \cosh({\mu k\over 2}) \sinh\left((\mu-\pi)\frac{k}{4}\right)}
- {\sinh\left(\frac{k}{2}\mu\right)\cos(k\mu b)
\over \cosh({\mu k\over 2}) \sinh\left((\mu-\pi)\frac{k}{2}\right)}\non \\
& & +{2\, \sinh({k\mu\over 2})\over \cosh({\pi k\over 2})\sinh\big((\pi-\mu){k\over 2}\big)}\sum_{j=1}^{{\nu-1\over 2}}\cosh({k\pi\over 2})\cos(k\,w_{j}^{(1)})
\Bigg\} \,.
\ee
The factor ${i\pi\over \nu-1}$ in (\ref{spinhalflatticeNLIE2}) is obtained as before by considering the $u\rightarrow\infty$ limit of (\ref{defa2}) and (\ref{spinhalflatticeNLIE2}). 

Next, taking the continuum limit ($\La \rightarrow \infty\,, N\rightarrow\infty\,, \Delta\rightarrow 0)$ and defining the renormalized rapidity $\theta$ as in (\ref{renormrapidity}),
(\ref{spinhalflatticeNLIE2}) becomes
\be
\ln \af(\theta) &=& 
\int_{-\infty}^{\infty}d\theta'\ \Gf (\theta-\theta'+i\varepsilon) \ln 
(1+\af(\theta'-i\varepsilon)) - 
\int_{-\infty}^{\infty}d\theta'\ \Gf(\theta-\theta'-i\varepsilon) \ln (1+\bar 
\af(\theta'+i\varepsilon))\non \\
&-& i 2mL \sinh \theta + i\, \Pf_{bdry}(\theta) +{i\pi\over \nu-1}\,,
\label{NLIE4}
\ee
where definitions (\ref{mathfrakdefs}) have been used.
By making following identification
\be
f(\theta) \equiv \ln (-\bar \af(\theta))
\ee
(\ref{NLIE4}) takes the following form,
\be
f(\theta) &=& 2i \int_{-\infty}^{\infty} d\theta'\ \Im m\
\Gf(\theta-\theta' - i \varepsilon)\
\ln (1 - e^{f(\theta' + i \varepsilon)})\non \\ 
&+& 2i m L \sinh \theta - i \Pf_{bdry}(\theta) + i\pi - {i\pi\over \nu-1}\,.
\label{nlie5}
\ee

\subsection{Boundary and Casimir energies}

Now we compute the boundary correction (order 1) and the Casimir correction (order $1/L$) to the energy.  
Using (\ref{energydef})-(\ref{deff2}) and (\ref{FTfact}), where now
\be
T^{(\pm)}(u)&=& \sinh(2u \pm i \mu)\, \sinh(u\pm{i\mu\over 2})\, \sinh(u\mp{3i\mu\over 2})\, \mathcal{\tilde{B}}^{(\pm)}(u)\, \phi(u\pm\frac{i\mu}{2})\non \\
&\times& \frac{Q(u\pm i\mu(\nu-1))}{Q(u)}
\label{Tpm2}
\ee
we have the following for the energy $E$,
\be
E&=&-\frac{g}{\Delta}\frac{d}{du}\left\{ \ln T^{(+)}(u+\frac{i\mu}{2})-
\ln T^{(-)}(u-\frac{i\mu}{2})\right\} \bigg\vert_{u=\Lambda} \non 
\\
&=&-\frac{g}{\Delta}\int \frac{dk}{2\pi} e^{-ik\Lambda}\left[
e^{\frac{\mu k}{2}}\widehat{LT^{(+)'}}(k)-e^{-\frac{\mu k}{2}}\widehat{LT^{(-)'}}(k)\right] \,,
\label{FTenergy}
\ee
where now one finds,
\be
e^{\frac{\mu k}{2}}\widehat{LT^{(+)'}}(k)&-&e^{-\frac{\mu k}{2}}\widehat{LT^{(-)'}}(k) \non \\
&=&
e^{{\mu k\over{2}}}
\widehat{L\mathcal{\tilde{B}}^{(+)'}}(k)+(e^{(\mu-\pi)k}-e^{-\mu k})\widehat{L\phi'}(k) 
+ (e^{(\mu-\frac{\pi}{2})k}-e^{-\mu k}) \frac{2\pi\psi_2(k)}{(-ik)}\non \\
&-& e^{-{\mu k\over{2}}}\widehat{L\mathcal{\tilde{B}}^{(-)'}}(k)
+ 4e^{-\frac{\pi k}{2}}\sinh((\pi-\mu)\frac{k}{2})\widehat{L\tilde{Q}'}(k) \non \\
&-& {16 i\pi\over\, \sinh(\pi k)}\sinh({k\mu\over 2})\sum_{j=1}^{{\nu-1\over 2}}\cosh({k\pi\over 2})\cos(k\,w_{j}^{(1)})\,,
\label{tpmdiff4}
\ee
$\widehat{LQ'}(k)$ is determined from (\ref{logQ2}) and $\widehat{LQ'}(k) = {1\over (-i k)}\widehat{LQ''}(k)$. As a result, after some algebra (\ref{tpmdiff4}) becomes,
\be
e^{\frac{\mu k}{2}}\widehat{LT^{(+)'}}(k)&-&e^{-\frac{\mu k}{2}}\widehat{LT^{(-)'}}(k) \non \\
&=& 2 \tanh ({\mu k\over 2})e^{-\frac{\pi k}{2}}\cosh\big((\mu-{\pi\over 2})k\big)\widehat{L\phi'}(k)\non \\
&+& \tanh({\mu k\over{2}})\bigg[-{2\pi i\over \sinh \frac{\mu k}{2}}
+ {2\pi i\cosh((\mu-\frac{\pi}{4})k)\over \sinh \frac{\pi k}{4}}\non \\
&+& {4\pi i\cosh((\mu-\frac{\pi}{2})k)\over \sinh \frac{\pi k}{2}} + e^{-{\mu k\over{2}}}\widehat{L\mathcal{\tilde{B}}^{(-)'}}(k)
+e^{{\mu k\over{2}}}\widehat{L\mathcal{\tilde{B}}^{(+)'}}(k)\bigg]\non \\
&+& \frac{1}{\cosh{\mu k\over{2}}}\left[\widehat{LA'}(k)+\widehat{{\cal L}{\bar A}'}(k)\right]
\label{tpmdiff5}
\ee
Note that the terms with extra-roots $\{w_{j}^{(1)}\}$ cancel out eventually. Introducing (\ref{tpmdiff5}) in (\ref{FTenergy}), one has the following result for the energy,
\beq
E = E_{L} + E_{1} + E_{1/L}
\label{FTenergyint}
\eeq
where
\be
E_{L} = -\frac{g}{2\pi \Delta}\int_{-\infty}^{\infty} dk\ e^{-ik\Lambda}\tanh({\mu k\over{2}})\left[2e^{-{\pi k\over 2}}\cosh((\mu-{\pi\over{2}})k) \widehat{L\phi'}(k)\right]
\label{bulkb} 
\ee
\be
E_{1} = -\frac{g}{2\pi \Delta}\int_{-\infty}^{\infty} dk\ e^{-ik\Lambda}\tanh({\mu k\over{2}})\bigg[&-&{2\pi i\over \sinh \frac{\mu k}{2}}
+ {2\pi i\cosh((\mu-\frac{\pi}{4})k)\over \sinh \frac{\pi k}{4}}\non \\
&+& {4\pi i\cosh((\mu-\frac{\pi}{2})k)\over \sinh \frac{\pi k}{2}}
+ e^{-{\mu k\over{2}}}\widehat{L\mathcal{\tilde{B}}^{(-)'}}(k)
+e^{{\mu k\over{2}}}\widehat{L\mathcal{\tilde{B}}^{(+)'}}(k)\bigg] \non \\
\label{boundb}
\ee
\be
E_{1/L} = -\frac{g}{2\pi \Delta}\int_{-\infty}^{\infty} dk\ e^{-ik\Lambda}\frac{1}{\cosh{\mu k\over{2}}}\left[
\widehat{LA'}(k)+\widehat{{\cal L}{\bar A}'}(k)\right]
\label{Casb} 
\ee
Equations (\ref{bulkb}), (\ref{boundb}) and (\ref{Casb}) represent the bulk, boundary and Casimir energies respectively.
The bulk energy $E_{L}$, is computed as before (refer to section 3.2) to yield
\beq
E_{L}=\frac{1}{4}Lm^2\cot\frac{\pi^2}{2\mu} \,,
\eeq
where $L$ and $m$ are defined by (\ref{continuumlimit}).
We now consider the boundary energy (\ref{boundb}). Using results (\ref{FT}) (with $\phi\rightarrow \mathcal{\tilde{B}}^{(\pm)}$) and (\ref{psi}), 
we have
\be
e^{-{\mu k\over{2}}}\widehat{L\mathcal{\tilde{B}}^{(-)'}}(k)+e^{{\mu 
k\over{2}}}\widehat{L\mathcal{\tilde{B}}^{(+)'}}(k)
&=&\frac{8i\pi}{\sinh \pi k}\cosh(\frac{\pi k}{2})\cos(k\mu b)
\label{bpbmdiff2}
\ee 
Substituting (\ref{bpbmdiff2}) into (\ref{boundb}), we obtain the boundary energy
\be
E_{1}={m\over 2}\left[1 + \cot \frac{\pi\nu}{4}  + 2\cot \frac{\pi\nu}{2}
 - 2\frac{\cosh\left(b\pi\right)}{\sin\frac{\pi\nu}{2}}\right] \,,
\label{FTbenergy3}
\ee
where we have used the same contour as for the bulk energy, namely by closing the contour in the lower half plane and selecting
only the contribution from the pole at $k=-\frac{i\pi}{\mu}$.  Comparing (\ref{FTbenergy3}) with Al. Zamolodchikov's result \cite{AZ1, BPT} for the energy of the
continuum sine-Gordon model (single boundary), 
\be
E(\eta \,, \vartheta) = -{m\over 2 \cos \left(\pi/ (2\lambda) \right)}
\left[-{1\over 2}\cos \left(\pi/ (2\lambda) \right) 
+{1\over 2}\sin \left(\pi/ (2\lambda) \right)-{1\over 2} 
+ \cos (\eta/\lambda) + \cosh (\vartheta/\lambda) \right] \,,
\label{AlZam}
\ee
and using $\lambda\equiv {8\pi\over \beta^{2}}-1 = {1\over \nu - 1}$, we conclude that
\be
\eta_{\pm} &=& (1 - \lambda){\pi\over 2}\,, \non \\ 
\vartheta_{\pm} &=& \lambda \pi b = (\lambda+1) \beta \,,
\label{boundparamreltn1}
\ee
One can verify that (\ref{boundparamreltn1}) coincides with the result in
\cite{AN} with $a_{\pm}=1$ (refer to equation (3.16) of that paper) for the constraint case (see equations (\ref{constraint}) and (\ref{realconstraints})) 
studied in section 3 of this paper. This suggests that the relation between the boundary parameters of the continuum model ($\eta_{\pm}\,, \vartheta_{\pm}$) and the lattice model ($a_{\pm}\,, b_{\pm}$)
in \cite{AN} might hold true in general. Consequently, the relation between the boundary parameters of the lattice model $(\alpha_{\pm}\,, \beta_{\pm})$
and that of the continuum action $(\mu_{\pm}\,, \varphi_{0}^{\pm})$ (see (\ref{SGboundaction})) as given in (\ref{boundparamreltn2}) should also be true for the case with arbitrary $\beta_{\pm}$,
hence possibly indicating that (\ref{boundparamreltn2}) might hold true in general as well.
One notes that for $\nu = 3\,, 5\,, 7\,, \ldots$, the boundary energy becomes
\be
E_{1}={m\over 2}\left[1 + (-1)^{{\nu-1\over 2}}(1 - 2\cosh\left(b\pi\right))\right] \,.
\label{FTbenergy4}
\ee 

Lastly, we consider the Casimir energy (\ref{Casb}). Passing to coordinate space and taking the continuum limit, we obtain
\be
E_C&=&\frac{2g}{i\Delta\mu}
\int_{-\infty}^{\infty}du\,\Im m \left(\frac{1}{\cosh\frac{\pi}{\mu}
(\Lambda-u-i\ep)}\right)'\ln (1+{\bar a}(u+i\ep))
\ee
which reduces to
\be
E_{C} = -\frac{m}{2\mu}
\int_{-\infty}^{\infty}du\, \Im m \, e^{\frac{\pi}{\mu}(u+i\ep)}
\ln (1+{\bar a}(u+i\ep))\,
\ee
in $\Lambda\rightarrow\infty$ limit (where $\left(\frac{1}{\cosh\frac{\pi}{\mu}(\Lambda-u-i\ep)}\right)'\rightarrow {2\pi\over \mu}e^{-{\pi\over \mu}(\Lambda-u-i\epsilon)}$) and after using
(\ref{continuumlimit}).
Similar manipulations of section 3.2 reduces $E_{C}$ to (We have used the fact that ${\bar \Af}(u)=\Af(-u)$ and $\Im m z=-\Im m{\bar z}$ in the process.) 
\be
E_{C} = -\frac{m}{2\pi}
\int_{-\infty}^{\infty}d\theta\, \Im m \, \sinh(\theta+i\varepsilon)\ln (1+{\bar \af}(\theta+i\varepsilon))\,
\ee
Invoking the identification $f(\theta) \equiv \ln(-{\bar \af}(\theta))$, the above becomes
\be
E_{C} = -{m\over 2\pi} \int_{-\infty}^{\infty} d\theta\ 
\Im m\ 
\sinh (\theta+ i \varepsilon)  \ln (1 - e^{f(\theta + i \varepsilon)})
\,.
\label{SGCasimir}
\ee

By considering the UV limit where $L\rightarrow 0$ and utilizing the steps in \cite{LMSS}, we obtain the following for the Casimir energy $E_{C}$,
\be
E_{C} = -{c\pi\over 24 L}
\ee
where the central charge $c$ is given by
\be
c &=& 1 - {6\over \pi^{2}}\Big({\nu-1\over \nu}\Big)(\Pf_{bdry}(\infty)+\pi)^{2}\non \\
&=& 1 - 6\Big({\nu\over \nu-1}\Big)
\label{c}
\ee
where $\Pf_{bdry}(\infty) = -\pi \big(2+{1\over \nu-1}\big)$.
In the following section, we present numerical results for the central charge of the corresponding open spin-$\frac{1}{2}$ XXZ spin chain, which agree with (\ref{c}).
We emphasize that for the case considered in section 3, the central charge of the sine-Gordon model (in the UV limit, $L\rightarrow 0$) was found to coincide
with that of the XXZ spin chain. See section 3.2.1 in \cite{AN} for more details.  

\subsection{Numerical results}

In this section, we present some numerical results and the extrapolated value for the central charge of XXZ spin chain. Such results are obtained as follows: 
We first numerically solve the Bethe equations (\ref{BAE2}) for some finite number of spins.  We use the solution to calculate Casimir energy from the following 
\be
E = E_{bulk} + E_{boundary} + E_{Casimir}
\label{casimirboundarybulk}
\ee
In (\ref{casimirboundarybulk}), $E$ is given by \cite{MN}
\be
E &=& {1\over 2} \sin^{2} \mu \sum_{k=1}^{M}{1\over 
\cosh (v_{k} - {i\mu\over2})
\cosh (v_{k} + {i\mu\over2})} + {1\over 2}(N-1) \cos \mu
\label{energyI2}
\ee 
Having determined the Bethe roots numerically, one uses known expressions for $E_{bulk}$ \cite{Yang}, namely
\be
E_{bulk} &=&  - N \sin^{2} \mu \int_{-\infty}^{\infty}
d\lambda\ {1\over \left[\cosh(2 \mu \lambda) - \cos \mu \right] 
\cosh (\pi \lambda)} +  {1\over 2}N \cos \mu \,,
\label{bulkenergy5}
\ee
and $E_{boundary}$ \cite{MN} given by 
\be
E_{boundary}&=& - {\sin \mu\over \mu} 
\int_{-\infty}^{\infty} d\omega\ 
{1\over 2\cosh (\omega/ 2)}
\Big\{ 
{\sinh((\nu-2)\omega/4) \over 2\sinh(\nu \omega/4)} 
-{1\over 2} \non \\
&+& {\sinh(\omega/2) \cosh((\nu-2)\omega/2)\over 
\sinh(\nu\omega/2)} + 
{\sinh(\omega/2) \cos (b\omega) \over 
\sinh(\nu \omega/2)} \Big\} -{1\over 2}\cos \mu \,.
\label{boundenergyeachI}
\ee
to determine $E_{Casimir}$. Then using $E_{Casimir} = -{\pi^{2}\sin\mu\over 24\mu\, N}c_{eff}$, one determines the effective central charge, $c_{eff}$ 
for that value of $N$,
\be
c_{eff} = -{24\mu N\over \pi^{2} \sin\mu}(E - E_{bulk} - E_{boundary})
\label{centralchargenumerical}
\ee
Since we are ultimately interested in $N\rightarrow\infty$ limit,  we employ an algorithm due to Vanden Broeck 
and Schwartz \cite{VBS}--\cite{Hamer3} to extrapolate these values for central charge to $N\rightarrow \infty$ limit.  
Table 1 below shows the $c_{eff}$ values for few finite even $N$, computed for some values of $\mu$ and $b$, for the lowest energy state that we considered in section 4.1 of this paper. The extrapolated
values (-8.02719 and -7.97641) obtained from the Vanden Broeck and Schwartz algorithm agree with the result obtained (= -8) from (\ref{c}). 

\begin{table}[htb] 
  \centering
  \begin{tabular}{|c|c|c|}\hline
    $N$ & $c_{eff}$, $b = 1.65$  & $c_{eff}$, $b = 1.79$\\
    \hline
    112      & -3.0586758038329473       & -2.510905382065773 \\ 
    120      & -3.1514592759743896      & -2.6001056254878296 \\
    128      & -3.2385273078233214      & -2.6842360822480313 \\
    136      & -3.3204955969641814      & -2.7638181748627413 \\
    144      & -3.3978856780639934      & -2.8392958325047744 \\
    152      & -3.4711433285531212      & -2.911050348946888 \\ 
    160      & -3.5406526531820894      & -2.9794118013959547 \\  
    168      & -3.6067470178385475      & -3.0446679508432566 \\
    176      & -3.6697176469097252      & -3.1070712683288697 \\
    184      & -3.7298204634077208      & -3.1668445456417547 \\
    192      & -3.787281586552167      & -3.2241854218624653 \\
    \vdots  &   \vdots        &  \vdots   \\
    extrapolated value            & -8.02719       & -7.97641\\ 
    \hline
   \end{tabular}
   \caption{Central charge $c_{eff}$, for $\nu = 3$ for two different boundary parameter values
    from numerical computations based on $N = 112$\,,$120$\,,\ldots\,,$192$ and 
    extrapolated values at $N\rightarrow \infty$ limit (Vanden Broeck and Schwartz algorithm).}
  \label{c2table}
\end{table}

The agreement between the calculated and the extrapolated values indicates that as with the constraint case studied in \cite{AN}, the central charge of the sine-Gordon
model (in UV limit), coincides with that of the corresponding spin-$\frac{1}{2}$ XXZ quantum spin chain. Furthermore, the numerical results also indicate that the extrapolated value of $c_{eff}$
is independent of the boundary parameter, as expected for models with Neumann boundary condition. 
This implies that the results for $c_{eff}$ and the conformal dimension $\Delta = {1-c_{eff}\over 24} = {\nu\over 4(\nu-1)}$ have more resemblance to spin chains with diagonal
boundary terms, as one would expect from the $\nu\over \nu-1$ dependance \cite{LMSS, ABR, Saleur, AOS} rather than $\nu-1\over\nu$ \cite{ABNPT} which is the anticipated
form for the conformal dimensions for spin chains with nondiagonal boundary terms. For more complete and detailed discussion on this, readers are urged to refer to \cite{Saleur, ABNPT}.
 
\section{Discussion}\label{sec:discuss}

From the proposed $T-Q$ equation of an open XXZ quantum spin chain with nondiagonal boundary terms, we 
derived the NLIE for the lowest energy state of an open spin-$\frac{1}{2}$ XXZ/sine-Gordon model with two boundaries. We first rederived the NLIE for
the case where the lattice boundary parameters obey certain constraint, which was treated in \cite{AN, ABNPT}. 
However, in contrast to the approach used there, we employed a method that utilizes the $T-Q$ equation of the spin chain model.
We next derived the NLIE for the case without such a constraint among the boundary parameters, where one of the lattice parameters
is set to be completely arbitrary. The lowest energy state of this spin chain model has complex sea of Bethe roots which is rather a feature common to critical spin-1 XXZ spin chain.
We also obtained the boundary energy and Casimir energy for the lowest energy state. We then presented relations between the boundary parameters of 
the continuum model and that of the lattice model which coincide with the ones found in \cite{AN} for the constraint case, hence suggesting
that these relations might hold true in general.  

Having found the $1/L$ correction, we proceeded to compute the central charge of the sine-Gordon model
in the UV limit. We also solved the corresponding spin-$\frac{1}{2}$ XXZ chain
numerically for some finite values of $N$. We used the solutions to compute $1/N$ correction
for these $N$ values, then extrapolate them to the $N\rightarrow\infty$ limit using Vanden Broeck 
and Schwartz algorithm. The extrapolated value of the effective central charge, $c_{eff}$ is found to be in agreement with that of the sine-Gordon model in UV limit.

In addition, the numerical results also indicate that $c_{eff}$
is independent of the boundary parameters, as expected for models with Neumann boundary condition. 
The result for the conformal dimension $\Delta$, turns out to be similar to that of the XXZ spin chain models with diagonal boundary terms
rather than the nondiagonal ones, to which the model studied here belongs to.  Such a feature however had been encountered before in literature \cite{M1}. There it was pointed out that
such a behaviour can be possibly attributed to spectral equivalences between diagonal-nondiagonal open XXZ spin chains \cite{GNPR}-\cite{Bajnok}.
 
There are other problems that one can explore and address further. For example, one could investigate the open spin-1 XXZ chain with general integrable boundary conditions 
in a similar manner, along the line of  \cite{ANS}, since solutions for such a model are already available \cite{FNR, M2}. It will also be interesting to study boundary excitations for these 
cases. We look forward to address these issues in near future.    

\section*{Acknowledgments}
I am grateful to the referee for pointing out a crucial reference.

\end{document}